\documentclass[preprint,12pt,a4paper]{elsarticle}

\makeatletter
\def\ps@pprintTitle{%
  \let\@oddhead\@empty
  \let\@evenhead\@empty
  \def\@oddfoot{\reset@font\hfil\thepage\hfil}
  \let\@evenfoot\@oddfoot
}
\makeatother

\usepackage{amssymb}

\usepackage{hyperref}
\usepackage{breakurl}

\usepackage[fleqn]{amsmath}
\begin{document}

\begin{frontmatter}




\title{A correction term for the asymptotic scaling of drag in flat-plate turbulent boundary layers}


\author[au1]{Nils T. Basse}
\ead{nils.basse@npb.dk}

\address[au1]{Trubadurens v\"ag 8, 423 41 Torslanda, Sweden \\ \vspace{10 mm} \small {\rm \today}}

%
%
%
%
%
%

\begin{abstract}
An asymptotic scaling law for drag in flat-plate turbulent boundary layers has been proposed [Dixit SA, Gupta A, Choudhary H, Singh AK and Prabhakaran T. Asymptotic scaling of drag in flat-plate turbulent boundary layers. Phys. Fluids {\bf 32}, 041702 (2020)]. In this paper we suggest to amend the scaling law by using a correction term derived from the logarithmic law for the mean velocity in the streamwise direction.
\end{abstract}

\end{frontmatter}



\section{Introduction}

In \cite{dixit_a}, an asymptotic (high Reynolds number) scaling law for drag in zero-pressure-gradient (ZPG) flow has been derived based on an approximation of $M$, the kinematic momentum rate through the boundary layer:

\begin{equation}
\label{eq:dixit_asymp}
  M = \int_0^{\delta} U^2 {\rm d}z \sim U_{\tau}^2 \delta,
\end{equation}

\noindent where $\delta$ is the boundary layer thickness, $U$ is the mean velocity in the streamwise direction, $z$ is the distance from the wall and $U_{\tau}$ is the friction velocity (we use $\sim$ to mean "scales as"). In this paper, we will propose a correction term to Equation (\ref{eq:dixit_asymp}).

The paper is structured as follows: In Section \ref{sec:derivation} we derive the correction term from the logarithmic law for the mean velocity in the streamwise direction. We apply this correction term to the measurements from \cite{dixit_a} in Section \ref{sec:application}, discuss the findings in Section \ref{sec:discussion} and conclude in Section \ref{sec:conclusions}.

\section{Derivation of the correction term}
\label{sec:derivation}

Our first step is to introduce the "log law" as formulated in \cite{marusic_a}:

\begin{equation}
\label{eq:log_law}
U^+ = \frac{1}{\kappa} \log (z^+) + A,
\end{equation}

\noindent where $U^+=U/U_{\tau}$, $z^+=z U_{\tau}/\nu$ is the normalized distance from the wall, $\nu$ is the kinematic viscosity, $\kappa$ is the von K\'arm\'an constant and $A$ is a constant for a given wall roughness. Although not strictly correct (close to and far from the wall), as our second step we will assume that the log law holds for the entire boundary layer of ZPG flows and use this to estimate the kinematic momentum rate through the boundary layer:

\begin{eqnarray}
\label{eq:M_log}
  M &=& \int_{0}^{\delta} U^2 {\rm d}z \nonumber \\
   &\sim& U_{\tau}^2 \delta \times \nonumber \\
   & & \left[ \frac{2}{\kappa^2} - \frac{2A}{\kappa} + A^2 +
\log (Re_{\tau}) \left( \frac{2A}{\kappa} - \frac{2}{\kappa^2} \right) + \log(Re_{\tau})^2/\kappa^2 \right],
\end{eqnarray}

\noindent where $Re_{\tau} = \delta U_{\tau} / \nu$ is the friction Reynolds number.

\begin{figure}[!ht]
\centering
\includegraphics[width=10cm]{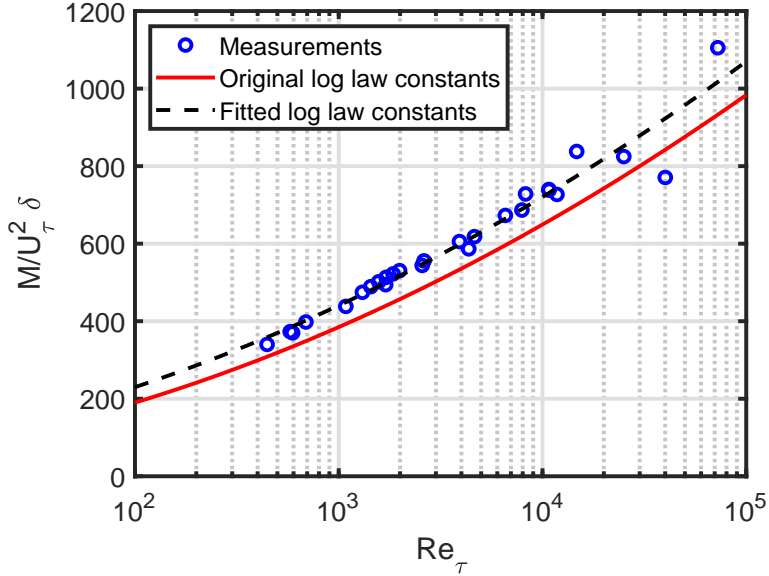}
\caption{$M/U_{\tau}^2 \delta$ as a function of $Re_{\tau}$. Blue circles are measurements from Table I in \cite{dixit_a}, the red solid (black dashed) line is the log law with original (fitted) constants, respectively.}
\label{fig:M}
\end{figure}

The term in the square brackets of Equation (\ref{eq:M_log}) is assumed to be a constant in Equation (\ref{eq:dixit_asymp}) \cite{dixit_a}; however, we show that it is in reality a function of $Re_{\tau}$. In Figure \ref{fig:M}, we show $M/U_{\tau}^2 \delta$ as a function of $Re_{\tau}$ using all measurements from Table I in \cite{dixit_a}. It is clear that this ratio varies with $Re_{\tau}$, i.e. it is not a constant and increases roughly a factor of 3 when $Re_{\tau}$ increases around two orders of magnitude. Also shown are two lines:

\begin{itemize}
  \item Red solid line: Log law with original constants from \cite{marusic_a}: $\kappa=0.39$ and $A=4.3$
  \item Black dashed line: Log law with fitted constants: $\kappa_{\rm fit}=0.39$ and $A_{\rm fit}=5.7$
\end{itemize}

Thus, we have demonstrated that $M$ is a function of both $\delta$ and $Re_{\tau}$:

\begin{equation}
\label{eq:basse_asymp}
M \sim U_{\tau}^2 \delta \times f(Re_{\tau}),
\end{equation}

\noindent where

\begin{equation}
f(Re_{\tau})=\left[ \frac{2}{\kappa^2} - \frac{2A}{\kappa} + A^2 +
\log (Re_{\tau}) \left( \frac{2A}{\kappa} - \frac{2}{\kappa^2} \right) + \log(Re_{\tau})^2/\kappa^2 \right]
\end{equation}

We note that the coefficient of determination $R^2$ with fitted constants is significantly larger than the one using the original constants, see Table \ref{tab:r_squared_log}. This shows that the fitted constants provide a better match than the original ones. However, we can not expect perfect agreement because of the assumptions made in deriving Equation (\ref{eq:M_log}).

\begin{table}[!ht]
\caption{Fit parameters and coefficient of determination ($R^2$) for the original and fitted $f(Re_{\tau})$.} 
\centering 
\begin{tabular}{cccc} 
\hline\hline 
Log law constants & $\kappa$ & $A$ & $R^2$ \\  
\hline 
Original & 0.39 & 4.3 & 0.80464 \\
Fitted   & 0.39 & 5.7 & 0.94960 \\
\hline 
\end{tabular}
\label{tab:r_squared_log} 
\end{table}

The asymptotic scaling law derived in \cite{dixit_a} is:

\begin{equation}
\label{eq:ZPG_asym_pow_law}
\tilde{U_{\tau}} \sim \frac{1}{\sqrt{\tilde{\delta}}},
\end{equation}

\noindent where

\begin{equation}
\tilde{U_{\tau}} = \frac{U_{\tau} \nu}{M} \sim \frac{\nu}{U_{\tau} \delta} = \frac{1}{Re_{\tau}}
\end{equation}

\noindent is named the "dimensionless drag" and

\begin{equation}
\tilde{\delta} = \frac{\delta M}{\nu^2} \sim \frac{\delta^2 U_{\tau}^2}{\nu^2} = Re_{\tau}^2
\end{equation}

\noindent scales as the friction Reynolds number squared.

Our conclusion is to propose that Equation (\ref{eq:basse_asymp}) should be used instead of Equation (\ref{eq:dixit_asymp}). As a consequence, Equation (\ref{eq:ZPG_asym_pow_law}) is modified to:

\begin{equation}
\tilde{U}_{\tau} \times \sqrt{f(Re_{\tau})} \sim \frac{1}{\sqrt{\tilde{\delta}}},
\end{equation}

\noindent where $\sqrt{f(Re_{\tau})}$ is the correction term.

\section{Application of the correction term}
\label{sec:application}

\begin{figure}[!ht]
\centering
\includegraphics[width=10cm]{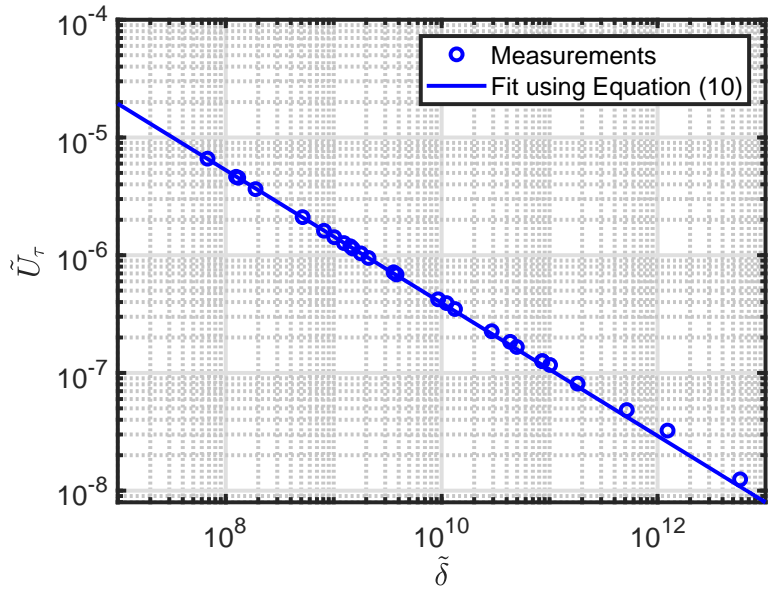}
\caption{Measurements from \cite{dixit_a} and fit to Equation (\ref{eq:p_law_dixit}).}
\label{fig:p_law_dixit}
\end{figure}

We fit all measurements in \cite{dixit_a} to:

\begin{equation}
\label{eq:p_law_dixit}
\tilde{U}_{\tau} = C \times \tilde{\delta}^D,
\end{equation}

\noindent where $C$ and $D$ are fit parameters, see Table \ref{tab:r_squared} and Figure \ref{fig:p_law_dixit}. Equations (7) and (8) in \cite{dixit_a} are both power-laws, but fitted to smaller and larger $\tilde{\delta}$ values, respectively: This is referred to as the "discrete model". Another model, the "continuous model" is presented as Equation (9) in \cite{dixit_a} and covers the entire range of $\tilde{\delta}$. As can be seen from Table \ref{tab:r_squared}, the $R^2$ of the continuous model is larger than the $R^2$ of the two discrete models, i.e. the continuous model performs better than the discrete models in fitting the measurements.

\begin{table}[!ht]
\caption{Fit parameters and coefficient of determination ($R^2$) for fits in \cite{dixit_a} and this paper.} 
\centering 
\begin{tabular}{cccc} 
\hline\hline 
Equation & $C$ & $D$ & $R^2$ \\  
\hline 
Equation (7) in \cite{dixit_a} & 0.15144 & -0.55745 & 0.99982 \\
Equation (8) in \cite{dixit_a} & 0.10869 & -0.54261 & 0.99992 \\
Equation (9) in \cite{dixit_a} & -       & -        & 0.99998 \\
Equation (10)                  & 0.17291 & -0.56439 & 0.99991 \\
Equation (11)                  & 1.06598 & -0.50629 & 0.99992 \\
Equation (12)                  & 1.23257 & -0.51017 & 0.99992 \\
\hline 
\end{tabular}
\label{tab:r_squared} 
\end{table}

\begin{figure}[!ht]
\centering
\includegraphics[width=10cm]{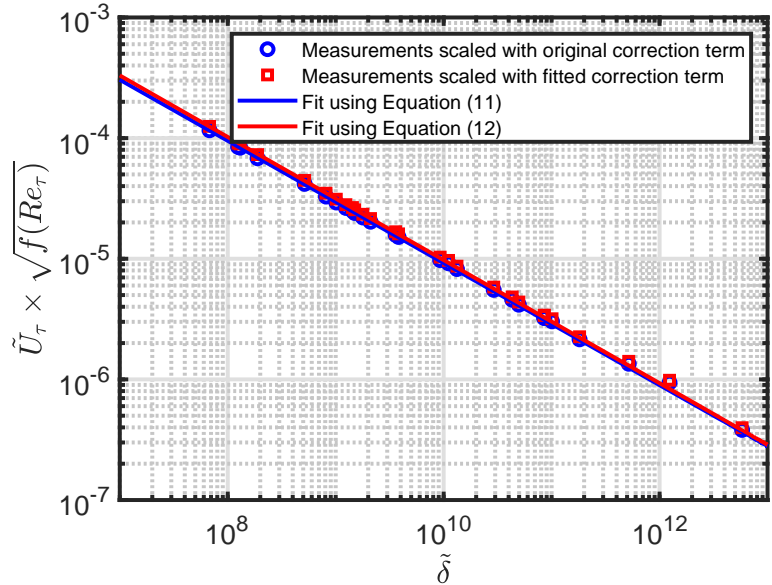}
\caption{Measurements from \cite{dixit_a} with correction terms applied and fits to Equations (\ref{eq:p_law_orig}) and (\ref{eq:p_law_fit}).}
\label{fig:p_law_mult}
\end{figure}

The next two fits are using the correction term $\sqrt{f(Re_{\tau})}$, either with the original log law constants:

\begin{equation}
\label{eq:p_law_orig}
\tilde{U}_{\tau} \times \sqrt{f(Re_{\tau})_{\rm original~constants}} = C \times \tilde{\delta}^D,
\end{equation}

\noindent or with the fitted log law constants:

\begin{equation}
\label{eq:p_law_fit}
\tilde{U}_{\tau} \times \sqrt{f(Re_{\tau})_{\rm fitted~constants}} = C \times \tilde{\delta}^D,
\end{equation}

\noindent see Table \ref{tab:r_squared} and Figure \ref{fig:p_law_mult}. The quality of the fits is similar to the one from Equation (\ref{eq:p_law_dixit}), but the fits with the correction term are interesting because their exponents are very close to 1/2. Thus, the deviation from 1/2 using Equation (\ref{eq:p_law_dixit}) may not only be because $Re$ is not sufficiently large, but also because the correction term is not included.

\section{Discussion}
\label{sec:discussion}

By comparing fit results from Equation (\ref{eq:p_law_dixit}) to results from Equations (\ref{eq:p_law_orig}) and (\ref{eq:p_law_fit}) - see Table \ref{tab:r_squared} - we find that the correction term scales weakly with $\tilde{\delta}$:

\begin{equation}
\sqrt{f(Re_{\tau})} \sim \tilde{\delta}^{0.05},
\end{equation}

\noindent which is the reason that the fits with the correction term have an exponent which is closer to 1/2.

For the case with correction term using the original log law constants (Equation (\ref{eq:p_law_orig})), we also see that the multiplier $C$ is close to 1 (1.06598, see Table \ref{tab:r_squared}); thus, for that case we propose an exact equation which matches the measurements quite well:

\begin{equation}
\tilde{U}_{\tau} \times \sqrt{f(Re_{\tau})_{\rm original~constants}} = \frac{1}{\sqrt{\tilde{\delta}}}
\end{equation}

Regarding measurements, we note that there is quite a large variation for large $Re_{\tau}$ (Figure \ref{fig:M}) and, equivalently, at high $\tilde{\delta}$ (Figures \ref{fig:p_law_dixit} and \ref{fig:p_law_mult}). This leads us to speculate that the measurements might have had different roughnesses, which e.g. impacts $A$ in the log law. It is not clear to us from the description in \cite{dixit_a} if this is indeed the case.

\section{Conclusions}
\label{sec:conclusions}

We have derived a correction term to the asymptotic scaling law of drag in ZPG turbulent boundary layers \cite{dixit_a}. The correction term has been applied to existing measurements and demonstrates that it leads to scaling with an exponent closer to -1/2 than the original scaling law.

\paragraph{Acknowledgements}

We are grateful to Google Scholar Alerts for making us aware of \cite{dixit_a} in a 'Recommended articles' e-mail dated 14th of May 2020.

\paragraph{Data availability statement}

Data sharing is not applicable to this article as no new data were created or analyzed in this study.


\label{sec:refs}

\end{document}